\documentclass{article}

\usepackage{graphicx}
\usepackage{natbib}
\usepackage{amsmath}
\usepackage{amssymb}

\usepackage[squaren]{SIunits}

\newcommand{\upi}{\pi}

\newcommand\St{\mbox{\textit{St}}}  % Strouhal number

\newcommand{\J}{\mathrm{j}}  % square root of -1
\newcommand{\D}{\mathrm{d}}  % differential
\newcommand{\E}{\mathrm{e}}  % exponential
\newcommand{\hankel}[1]{\mbox{H}_{#1}^{(1)}}
\newcommand{\besselj}[1]{\mbox{J}_{#1}}

\newcommand{\mathsfbi}[1]{\mathsf{#1}}

\newsavebox{\astrutbox}
\sbox{\astrutbox}{\rule[-5pt]{0pt}{20pt}}

\author{Michael Carley}
\title{Radial cancellation in spinning sound fields}

\begin{document}

\maketitle

\begin{abstract}
%   This paper considers the problem of sound radiated by a disc source
%   with sinusoidal azimuthal variation and arbitrary radial
%   variation. On the basis of an exact analysis, without recourse to
%   far field approximations, the number of degrees of freedom in a
%   field of azimuthal order $n$ is found to be less than $k-n$, where
%   $k$ is the non-dimensional wavenumber. The implications of this
%   result are discussed for the case of noise from rotors and from
%   turbulent jets.
  The radiating part of a circular acoustic source is determined, on
  the basis of an exact analysis of the radiation properties of a
  source with angular dependence $\exp\J n\theta$ and arbitrary radial
  dependence. It is found that the number of degrees of freedom in the
  radiated field is no greater than $k-n$, where $k$ is the
  wavenumber. The radiating part of the source at low frequency is
  explicitly stated and used to analyze noise cancellation. The
  results are applied to the identification of sources in jet noise
  and an explanation for the low order structure of jet noise fields
  is proposed.
\end{abstract}

% \begin{keywords}
%   Acoustics, noise, rotating sources, spinning sound fields, rotors,
%   fans, propellers, jets
% \end{keywords}

\section{Introduction}
\label{sec:intro}

Sound generation by spinning modes is a central problem in many
applications. Devices such as propellers and fans obviously produce a
rotating source system, but the termination of a circular duct also
radiates like a spinning source and jets have a source system which
can be decomposed into spinning modes. The problem of the relationship
between a source distribution and its spinning acoustic field has thus
attracted considerable attention in the literature.

This paper examines one part of the problem, the relationship between
the radial structure of the source and the form of the corresponding
acoustic field. Earlier work, using exact and asymptotic analysis
\citep{carley10b,carley10c}, has fixed an upper limit on the
information which is radiated into the field, but without considering
the effect of the source distribution. In this paper, the theory is
extended to include the relationship between the source and the field,
allowing a discussion of the implications for a number of problems.

A first area where the analysis is relevant is that of control of
noise from rotors. One approach to this problem is to fit an inverse
model to the measured noise and then use this model to compute a noise
field which cancels the rotor noise at some point. In a recent study
\cite[][]{gerard-berry-masson05a,gerard-berry-masson05b,%
  gerard-berry-masson-gervais07} it was found that a very low order
acoustic model of a cooling fan was sufficiently accurate for control
purposes. At first glance, it is not clear why a low order model
should give a good match to the results from a finite rotor, beyond
considerations of acoustical compactness. An analysis of the
information content of the field, however, shows that the field is
generated by a set of low order modes with the higher order modes
being cut-off and generating exponentially small noise
\cite[][]{carley10b,carley10c}.

The second application of the approach of this paper is to inverse
methods, in which a `source' is determined from acoustic field
measurements. There have been numerous applications of such
techniques, but of particular interest here is that of jet noise. It
is known that the noise field of a turbulent jet is represented by a
much lower order model than is the flow
\citep{jordan-schlegel-stalnov-noack-tinney07}. This is partly
explained by axial interference effects \citep[for example]{freund01}
but relatively little attention has been paid to the influence of the
radial structure of the source until quite recently \citep{michel09}. 

This paper presents an analysis of the general problem of radiation
from a disc source and fixes limits on the proportion of the source
which actually radiates into the acoustic field, with no approximation
other than the standard acoustic assumption of linearity. In
particular, the far-field assumption is not required, making the
results applicable over a wide range of parameters. The implications
of the results are then discussed with respect to rotor and jet
noise.

\section{Spinning sound fields}
\label{sec:spin}

\begin{figure}
  \centering
  \centerline{\includegraphics{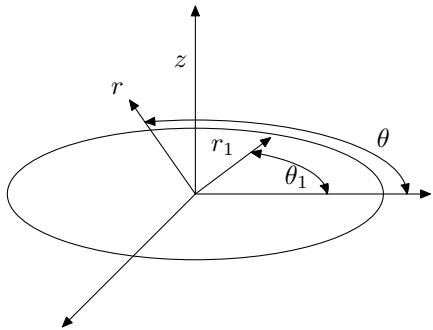}}
  \caption{Coordinate system for disc radiation calculations}
  \label{fig:coordinate}
\end{figure}

The problem is formulated as that of calculating the acoustic field
radiated by a monopole source distributed over a circular disc. This
disc may be viewed as the source proper, such as in the case of rotor
noise, or as part of a distributed three-dimensional source, as in jet
noise. The system for the analysis is shown in
Figure~\ref{fig:coordinate} with cylindrical coordinates
$(r,\theta,z)$. All lengths are non-dimensionalized on disc radius and
the disc lies in the plane $z=0$. The field from one azimuthal mode of
the acoustic source, specified as $s_{n}(r_{1})\exp\J n \theta_{1}$,
has the form $P_{n}(k,r,z)\exp\J n \theta$, with $P_{n}$ given by the
Rayleigh integral:
\begin{eqnarray}
  \label{equ:disc}
  P_{n}(k,r,z) &=& 
  \int_{0}^{1}
  \int_{0}^{2\upi} \frac{\E^{\J(kR'+n\theta_{1})}}{4\upi
    R'}\,\D  \theta_{1} s_{n}(r_{1}) r_{1}\,\D r_{1},\\
  R' &=& 
  \left[
    r^{2} + r_{1}^{2} - 2rr_{1}\cos\theta_{1} + z^{2}
  \right]^{1/2},\nonumber
\end{eqnarray}
where $k$ is wavenumber and subscript $1$ indicates variables of
integration. The field due to higher order sources, such as dipoles
and quadrupoles, would be found by differentiation of
(\ref{equ:disc}). The analysis to be presented does not include such
sources but the conclusions drawn should still be valid.

\subsection{Equivalent line source expansion}
\label{sec:line}

\begin{figure}
  \centering
  \centerline{\includegraphics{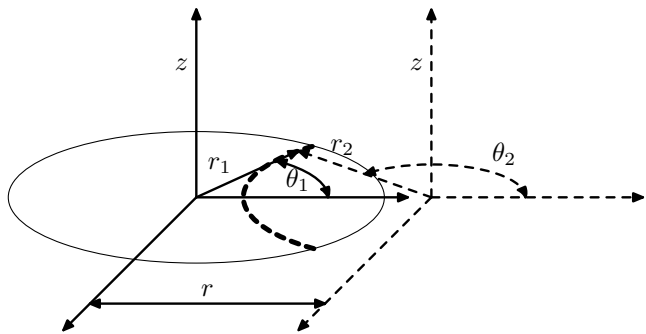}}
  \caption{Transformation to equivalent line source}
  \label{fig:sideline}
\end{figure}

The analysis of the nature of the sound field from an arbitrary disc
source is based on a transformation of the disc to an exactly
equivalent line source, an approach which has been used to study
transient radiation from pistons \cite[][]{oberhettinger61a,pierce89},
rotor noise \cite[][]{chapman93,carley99} and source identification
methods \citep{carley09}. 

% The method has also been recently used in a
% more detailed analysis of the radiation properties of disc sources
% \cite[][]{carley10b,carley10c}. In this paper, we do not need such
% detail in order to establish which parts of the source radiate and
% which do not.

The transformation to a line source is shown in
Figure~\ref{fig:sideline}, which shows the new coordinate system
$(r_{2},\theta_{2},z)$ centred on a sideline of constant radius
$r$. Under this transformation:
\begin{eqnarray}
  \label{equ:transformed}
  P_{n}(r,z) &=& \int_{r-1}^{r+1} \frac{\E^{\J kR'}}{R'}
  K(r,r_{2})r_{2}\,\D
  r_{2},\\
  R' &=& \left(r_{2}^{2} + z^{2}\right)^{1/2},\nonumber\\
  \label{equ:kfunc}
  K(r,r_{2}) &=& \frac{1}{4\upi}
  \int_{\theta_{2}^{(0)}}^{2\upi-\theta_{2}^{(0)}} \E^{\J
    n\theta_{1}}s_{n}(r_{1})\,\D\theta_{2},
\end{eqnarray}
for observer positions with $r>1$, with the limits of integration
given by:
\begin{equation}
  \label{equ:theta}
    \displaystyle
    \theta_{2}^{(0)} = \cos^{-1}\frac{1-r^{2}-r_{2}^{2}}{2rr_{2}}.
\end{equation}
Functions of the form of $K(r,r_{2})$ have been analyzed in previous
work \citep{carley99}. In this paper, it is sufficient to note that
the source function can be written \citep{carley09}:
\begin{eqnarray}
  \label{equ:kfunc:exp}
  K(r,r_{2}) &=& \sum_{q=0}^{\infty} u_{q}(r)U_{q}(s)(1-s^2)^{1/2},
\end{eqnarray}
where $U_{q}(s)$ is a Chebyshev polynomial of the second kind,
$s=r_{2}-r$ and the coefficients $u_{q}(r)$ are functions of $r$ but
not of $z$. Inserting (\ref{equ:kfunc:exp}) into
(\ref{equ:transformed}):
\begin{eqnarray}
  \label{equ:line:ip}
  P_{n}(k,r,z) &=&
  \sum_{q=0}^{\infty} u_{q}(r)
  \int_{-1}^{1} \frac{\E^{\J kR'}}{R'}
  U_{q}(s) (r+s)(1-s^2)^{1/2}\,\D s,\\
  R' &=& \left[(r+s)^{2} + z^{2}\right]^{1/2}.
\end{eqnarray}
The radiation properties of the integrals of (\ref{equ:line:ip}) have
been examined in some detail elsewhere
\citep[][]{carley10b,carley10c}, giving an exact result for the
in-plane case $z=0$:
\begin{equation}
  \label{equ:line:series}
  P_{n}(k,r,0) =
  \upi \E^{\J k r} \sum_{q=0}^{\infty} u_{q}(r)
  \J^{q} (q+1) \frac{\besselj{q+1}(k)}{k}.
\end{equation}
For large order $q$, the Bessel function $J_{q}(k)$ decays
exponentially for $k<q$ so that the line source modes with order $q>k$
are `cut-off' and generate exponentially small noise fields. Since the
integrals have their maximum in the plane $z=0$,
(\ref{equ:line:series}) says that the whole field is exponentially
small and modes $u_{q}(r)$ with $q>k$ are cut-off everywhere. This
gives an indication of how much of a given source distribution
radiates into the acoustic field, near or far. There only remains to
establish the relationship between the radial source $s_{n}(r_{1})$
and the line source coefficients $u_{q}(r)$.

\subsection{Series expansion for spinning sound fields}
\label{sec:series}

A recently derived series \citep{carley10} for the field radiated by a
ring source of radius $r_{1}$ can be used to find an expression for
the sound radiated by a disc source with arbitrary radial variation:
\begin{eqnarray}
  R_{n} &=& \int_{0}^{2\upi} \frac{\E^{\J(kR'+n\theta_{1})}}{4\upi
    R'}\,\D 
  \theta_{1},
  \nonumber\\
  &=&
  \J^{2n+1}\frac{\upi}{4}
  \frac{1}{(r_{1}R)^{1/2}}
  \sum_{m=0}^{\infty}
  (-1)^m
  \frac{(2n+4m+1)(2m-1)!!}{(2n+2m)!!}\nonumber\\
  \label{equ:ring}
  &\times&
  \hankel{n+2m+1/2}(kR)  P_{n+2m}^{n}(\cos\phi)
  \besselj{n+2m+1/2}(kr_{1}),
\end{eqnarray}
with $\hankel{\nu}(x)$ the Hankel function of the first kind of order
$\nu$, $\besselj{\nu}$ the Bessel function of the first kind and
$P_{n}^{m}$ the associated Legendre function. The observer position is
specified in spherical polar coordinates $R=[r^{2}+z^{2}]^{1/2}$,
$\phi=\tan^{-1}r/z$.

Multiplication by the radial source term $r_{1}s_{n}(r_{1})$ and
integration gives an expression for the field radiated by a general
source of unit radius and azimuthal order $n$:
\begin{eqnarray*}
  P_{n}(k,r,z) &=&
  \J^{2n+1}\frac{\upi}{4}
  \sum_{m=0}^{\infty}
  (-1)^m
  \frac{(2n+4m+1)(2m-1)!!}{(2n+2m)!!} P_{n+2m}^{n}(\cos\phi)S_{n+2m},\\
  S_{n+2m}(k,r,z) &=& \int_{0}^{1} s_{n}(r_{1})
  \besselj{n+2m+1/2}(kr_{1})\hankel{n+2m+1/2}(kR)  
  \left(
    \frac{r_{1}}{R}
  \right)^{1/2}\,\D r_{1}.
\end{eqnarray*}

Setting $z=0$ ($\phi=\upi/2$, $R=r$):
\begin{eqnarray}
  \label{equ:series:ip}
  P_{n}(k,r,0) &=&
  \frac{\J\upi}{4}
  \sum_{m=0}^{\infty}
  \frac{1}{m!}
  \frac{(2n+4m+1)(2n+2m-1)!!(2m-1)!!}{2^{m} (2n+2m)!!}
  S_{n+2m},
\end{eqnarray}
where use has been made of the
expression~\citep[8.756.1]{gradshteyn-ryzhik80}:
\begin{equation}
  P_{n+2m}^{n}(0) = \frac{(-1)^{m+n}}{2^{m}}\frac{(2n+2m-1)!!}{m!}.
\end{equation}

% The expansion of (\ref{equ:series:ip}) is exact and is equal to the
% exact expansion of (\ref{equ:line:series}). Equating the two 

\subsection{Line source coefficients}
\label{sec:coefficients}

The expressions for $P_{n}$ from \S\,\ref{sec:series}
and~\S\,\ref{sec:line} are both exact and can be equated to derive a
system of equations relating the coefficients $u_{q}(r)$ to the radial
source distribution $s_{n}(r_{1})$:
\begin{eqnarray}
  \frac{\J}{4}
  \sum_{m=0}^{\infty}
  \frac{1}{m!}
  \frac{(2n+4m+1)(2n+2m-1)!!(2m-1)!!}{2^{m}(2n+2m)!!}
  S_{n+2m}
  &=&\nonumber\\
  \label{equ:system}
  \E^{\J k r}
  \sum_{q=0}^{\infty} u_{q}(r)
  \J^{q} (q+1) \frac{\besselj{q+1}(k)}{k}.
\end{eqnarray}
Under repeated differentiation, (\ref{equ:system}) becomes a lower
triangular system of linear equations which connects the coefficients
$u_{q}(r)$ and $S_{n+2m}$:
\begin{eqnarray}
  \frac{\J}{4}
  \sum_{m=0}^{\infty}
  \frac{1}{m!}
  \frac{(2n+4m+1)(2n+2m-1)!!(2m-1)!!}{2^{m}(2n+2m)!!}
  S_{n+2m}^{(v)} &=&\nonumber\\
  \label{equ:diff}
  \sum_{q=0}^{\infty} u_{q}(r)
  \J^{q} (q+1)
  \left[
    \E^{\J k r}
    \frac{\besselj{q+1}(k)}{k}
  \right]^{(v)},
\end{eqnarray}
where superscript $(v)$ denotes the $v$th partial derivative with
respect to $k$, evaluated at $k=0$.

Using standard series \citep{gradshteyn-ryzhik80}, the products of
special functions can be written:
\begin{eqnarray}
  \label{equ:prod:exp:j}
    \E^{\J k r}\frac{\besselj{q+1}(k)}{k} &=&
    \frac{1}{\J^{q}} \sum_{t=0}^{\infty}(\J k)^{t+q} E_{t,q}(r),\\
    E_{t,q}(r) &=&  \frac{1}{2^{q+1}}
    \sum_{s=0}^{[t/2]}
    \frac{r^{t-2s}}{4^{s}s!(s+q+1)!(t-2s)!},\nonumber
\end{eqnarray}
where $[t/2]$ is the largest integer less than or equal to $t/2$, and
\begin{eqnarray}
  \left(
    \frac{r_{1}}{r}
  \right)^{1/2}
  \hankel{n+1/2}(kr)\besselj{n+1/2}(kr_{1}) &=& 
  \left(
    \frac{r}{2}
  \right)^{2n+1}
  \sum_{t=0}^{\infty}
  \frac{k^{2t+2n+1}}{t!}
  \left(
    -\frac{r^{2}}{4}
  \right)^{t}V_{n,t}(r_{1}/r) \nonumber\\
  &-& (-1)^n\J \sum_{t=0}^{\infty}
  \frac{k^{2t}}{t!}
  \left(
    -\frac{r^{2}}{4}
  \right)^{t}W_{n,t}(r_{1}/r),
  \label{equ:prod:h:j}
\end{eqnarray}
with the polynomials $V_{n,t}$ and $W_{n,t}$ given by:
\begin{subequations}
  \label{equ:vwpoly}
  \begin{eqnarray}
    V_{n,t}(x) &=& \sum_{s=0}^{t} {t \choose s}
    \frac{x^{2s+n+1}}{\Gamma(n+s+3/2)\Gamma(t-s+n+3/2)},\\
    W_{n,t}(x) &=& \sum_{s=0}^{t} {t \choose s}
    \frac{x^{2s+n+1}}{\Gamma(n+s+3/2)\Gamma(t-s-n+1/2)}.
  \end{eqnarray}
\end{subequations}

Given the power series, the derivatives at $k=0$ are readily found:
\begin{subequations}
  \label{equ:derivatives}
  \begin{eqnarray}
    \J^{q}\left.
      \frac{\partial^{v}}{\partial k^{v}}
      \E^{\J k r}\frac{\besselj{q+1}(k)}{k}
    \right|_{k=0}
    &=& 
    \left\{
    \begin{array}{ll}
      0, & v < q;\\
      \J^{v}v!E_{v-q,q}(r), & v \geq q.
    \end{array}
    \right.\\
     \left.
      \frac{\partial^{v}}{\partial k^{v}}
      (r_{1}/r)^{1/2}
       \hankel{n+1/2}(kr)\besselj{n+1/2}(kr_{1})
     \right|_{k=0}
     &=& \nonumber\\
     \left\{
     \begin{array}{lll}
       \displaystyle
       0, & v=2v'+1,& v' < n;\\
       \displaystyle
       \left(
         \frac{r}{2}
       \right)^{2n+1}
       \left(
         -\frac{r^{2}}{4}
       \right)^{v'-n}
       \frac{v!}{(v'-n)!}V_{n,v'-n}(r_{1}/r), & v=2v'+1, & v'\geq n;\\
       \displaystyle
       -(-1)^{n}\J \frac{(2v')!}{v'!}
       \left(
         -\frac{r^{2}}{4}
       \right)^{v'}W_{n,v'}(r_{1}/r), &v=2v'.
     \end{array}
     \right.
  \end{eqnarray}
\end{subequations}
Setting $v=0,1,\ldots$ yields an infinite lower triangular system of
equations for $u_{q}(r)$:
\begin{equation}
  \label{equ:system:1}
  \mathsfbi{E}\mathbf{U} = \mathbf{B},
\end{equation}
with $\mathbf{U}=[u_{0}\,u_{1}\,\ldots]^{T}$ and the elements of
matrix $\mathsfbi{E}$ and vector $\mathbf{B}$ given by:
\begin{subequations}
  \label{equ:entries}
  \begin{eqnarray}
    E_{vq} &=& 
    \begin{array}{ll}
      \J^{v}(q+1)v!E_{v-q,q}(r), & q\leq v;\\
      0, & q > v.
    \end{array}\\
    B_{v} &=&   \frac{\J}{4}
    \sum_{m=0}^{\infty}
    \frac{1}{m!}
    \frac{(2n+4m+1)(2n+2m-1)!!(2m-1)!!}{2^{m}(2n+2m)!!}
    S_{n+2m}^{(v)},
  \end{eqnarray}
\end{subequations}
where
\begin{eqnarray}
  \label{equ:integrals}
  S_{n+2m}^{(v)} &= 
  \left\{
    \begin{array}{lll}
      \displaystyle
      0 & v = 2v'+1, & v' < n+2m;\\
      \displaystyle
      (-1)^{n+v'}
      \frac{v!}{(v'-n-2m)!} 
      \left(
        \frac{r}{2}
      \right)^{v}
      \int_{0}^{1}V_{n+2m,v'-n-2m}(r_{1}/r) s_{n}(r_{1})\,\D r_{1}, 
      & v = 2v'+1, & v' \geq n+2m;\\
      \displaystyle
      -(-1)^{n+v'} \frac{\J v!}{v'!} 
      \left(
        \frac{r}{2}
      \right)^{v}
      \int_{0}^{1}W_{n+2m,v'}(r_{1}/r) s_{n}(r_{1})\,\D r_{1}, & v =
      2v'.
    \end{array}
  \right.
\end{eqnarray}
Given a radial source term $s_{n}(r_{1})$, (\ref{equ:system:1}) can be
solved to find the coefficients $u_{q}(r)$ of the equivalent line
source modes. Since it is lower triangular, the first few values of
$u_{q}$ can be reliably estimated, although ill-conditioning prevents
accurate solution for arbitrary large $q$.

% The system is poorly conditioned, however, and does not
% give reliable results for large $q$. On the other hand, the system is
% lower triangular meaning that the first few values of $u_{q}$ can be
% determined accurately by solution of (\ref{equ:system:1}) and certain
% properties of the coefficients can be derived analytically.

\subsection{Radiated field}
\label{sec:radiated}

From the relationship between the radial source term and the line
source coefficients, some general properties of the acoustic field can
be stated. The first result, already found by
\cite{carley10b,carley10c} is that, since the line source modes with
$q+1>k$ are cut off, the acoustic field has no more than $k$ degrees
of freedom, in the sense that the radiated field is given by a
weighted sum of the fields due to no more than $k$ elementary
sources. From (\ref{equ:system:1}), this result can be extended. 

The first extension comes from the fact that $B_{2v+1}\equiv0$, for
$v'<n$, on the right hand side of (\ref{equ:system:1}). This means
that $u_{q}$, $q=2v'+1$, is uniquely defined by the lower order
coefficients with $q\leq 2v'$. The result is that the acoustic field
of azimuthal order $n$, whatever might be its radial structure, has no
more than $k-n$ degrees of freedom.

A second extension comes from examination of (\ref{equ:system:1}) The
first few entries of the system of equations are:
\begin{equation}
  \label{equ:system:a}
  \left[
    \begin{array}{rrrrr}
      1/2 & 0 & 0 & 0 & \cdots \\
      r/2 & 1/4 & 0 & 0 & \cdots \\
      \vdots &\vdots & \vdots & 0 & \cdots
    \end{array}
  \right]
  \left(
    \begin{array}{c}
      u_{0} \\ u_{1} \\ \vdots
    \end{array}
  \right)
  =
  \left(
    \begin{array}{c}
      B_{0} \\ 0 \\ \vdots
    \end{array}
  \right),
\end{equation}
resulting in the solution:
\begin{equation}
  \label{equ:system:sol}
  u_{0} = 2B_{0};\quad u_{1} = -2ru_{0} = -4rB_{0},
\end{equation}
so that the ratio of $u_{0}$ and $u_{1}$ is constant, for arbitrary
$s_{n}(r_{1})$. This means that low frequency sources of the same
radius and azimuthal order generate fields which vary only by a
scaling factor, since the higher order terms are cut off.

Finally, if we attempt to isolate a source $s_{n}(r_{1})$ associated
with a single line source mode, by setting $u_{q}\equiv1$ for some
$q$, with all other $u_{q}\equiv0$, we find that the line modes must
occur in pairs, since if $u_{2v'}\equiv1$, $u_{2v'+1}\neq0$, being
fixed by the condition $B_{2v'+1}\equiv0$.

\section{Results}
\label{sec:results}

To illustrate the application of the result of the previous section,
we present some results for the calculation of the line source
coefficients and for the use of the method to modify the radiating
part of a source. We also discuss qualitatively the implications of
the results for studies of jet noise.

\subsection{Line source coefficient evaluation}
\label{sec:coefficient}

\begin{figure}
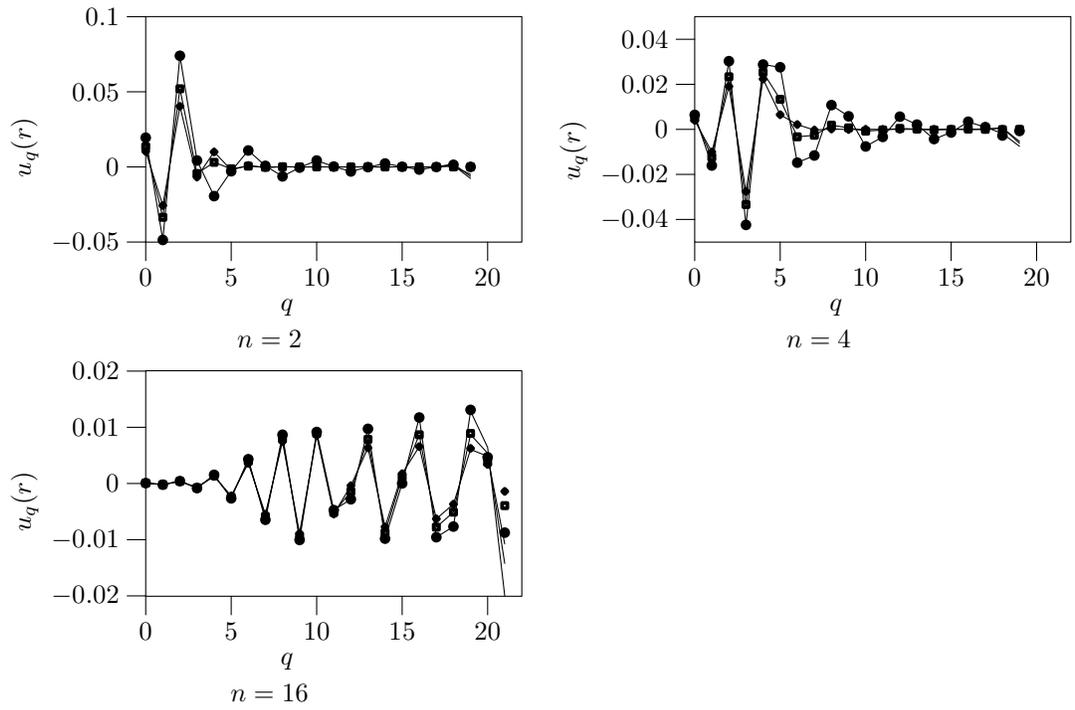

  \centering
  \begin{tabular}{cc}
    \includegraphics{jfm10-figs.3} &
    \includegraphics{jfm10-figs.4} \\
    $n=2$ & $n=4$ \\
    \includegraphics{jfm10-figs.5} \\
    $n=16$
  \end{tabular}
%   \centerline{\includegraphics{jfm10-figs.3}}
%   \centerline{\includegraphics{jfm10-figs.4}}
%   \centerline{\includegraphics{jfm10-figs.5}}
  \caption{Line source mode coefficients computed using the method of
    \S\,\ref{sec:coefficients} (solid lines) and directly from
    analytical formulae (symbols) for $r=5/4$, $s=r_{1}^{a}$, $a=0$
    (circles), $a=2$ (squares) and $a=4$ (diamonds) for $n=2,4,16$.}
  \label{fig:cfft:compare}
\end{figure}

The first results are a comparison of coefficients $u_{q}(r)$ computed
using (\ref{equ:system:1}) and those computed directly from exact
closed-form expressions for $K(r,r_{2})$ in the case when the radial
source term is a monomial in radius $s_{n}=r_{1}^{a}$
\citep{carley99}. Figure~\ref{fig:cfft:compare} compares the two sets
of coefficients for $a=0,2,4$, with the plots terminated at a value of
$q$ where the difference between the two sets of results becomes
noticeable, $q\approx20$. This gives an indication of the effect of
the ill conditioning of (\ref{equ:system:1}). For $q\lesssim 20$, the
computed values of $u_{q}$ are reliable. It is noteworthy that for
small $q$, the coefficients are practically equal for all values of
$a$ so that for low frequency radiation, the radiated fields will be
practically indistinguishable.

% As a first check on the result of \S\,\ref{sec:coefficients}, the
% coefficients $u_{q}(r)$ for analytically specified sources $s(r_{1})$
% were computed. Exact closed-form expressions for $K(r,r_{2})$ have
% been published previously for the case when the source is a power of
% radius, $s_{n}=r_{1}^a$
% \citep{carley99}. Figure~\ref{fig:cfft:compare} shows the coefficients
% $u_{q}$ computed using the method of \S\,\ref{sec:coefficients},
% compared to those found directly from the analytical expression for
% $K(r,r_{2})$. The plots are terminated at the point where the
% difference between $u_{q}$ computed using the two methods is
% noticeable, giving an indication of how the ill-conditioning of the
% system of equations in \S\,\ref{sec:coefficients} affects the
% result. For the data presented here, which are representative of all
% of the configurations examined, this happens at $q\approx20$. Below
% this value, the calculated values of $u_{q}$ are reliable.

% A second point worthy of note is that the coefficient values for small
% $q$ are practically equal for different values of $a$. This is
% especially clear for the $n=16$ results, where the difference between
% coefficients only becomes obvious for $q\gtrsim15$. For low frequency
% radiation, $k<15$, the higher order modes will be cut off and the
% sound fields from the different source terms will be indistinguishable
% since only the lower order modes, which have the same coefficients,
% radiate.

\subsection{Radial cancellation}
\label{sec:cancellation}

\begin{figure}
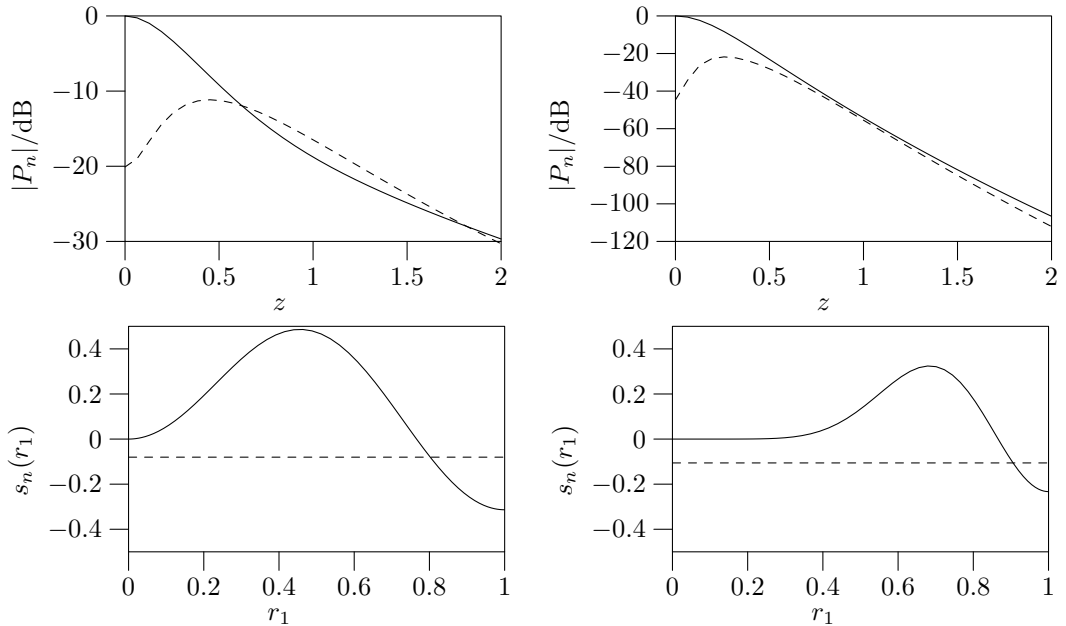

  \centering
  \begin{tabular}{cc}
    \includegraphics{jfm10-figs.6} &
    \includegraphics{jfm10-figs.8} \\
    \includegraphics{jfm10-figs.7} &
    \includegraphics{jfm10-figs.9} 
  \end{tabular}
  \caption{Cancellation effects for radial source terms: top row
    radiated field from original $s_{n}(r_{1})$ (solid) and modified
    source $s_{n}(r_{1})-\beta s_{n}'(r_{1})$ (dashed); bottom row source
    terms $s_{n}(r_{1})$ (solid) and $\beta s_{n}'(r_{1})$ dashed. The left
    hand column covers the case $n=2$ and the right hand $n=8$. In
    both cases, $k=1$, $r=5/4$.}
  \label{fig:field:compare}
\end{figure}

The analysis so far has identified that part of a source distribution
which radiates, as a function of wavenumber $k$. From the results, it
appears that only a small part of the source is responsible for the
acoustic field, with most of the line source modes being cut off
throughout the field, except at high frequency. Indeed, asymptotic
analysis \citep{carley10b,carley10c} shows that even the cut-on modes
radiate efficiently only into a small part of the acoustic field, with
the exception of those of low order. In any case, this offers a method
for examination of the radiation properties of a source. Given a
source term $s_{n}(r_{1})$ the approach is to impose a secondary
source $s_{n}'(r_{1})$ which generates the same set of line source
modes up to some required order. In the simplest case, we match the
first line source coefficient $u_{0}$, which automatically matches
$u_{1}$. If $u_{0}$ is known for both $s_{n}(r_{1})$ and for
$s_{n}'(r_{1})$, then the source $s_{n}'(r_{1})-\beta s_{n}'(r_{1})$
will have $u_{0}=u_{1}=0$, if $\beta$ is taken as the ratio of $u_{0}$
for the primary and secondary sources $s_{n}(r_{1})$ and
$s_{n}'(r_{1})$.

Figure~\ref{fig:field:compare} shows the results of such a procedure
using sources $s_{n}(r_{1})=J_{n}(\alpha r_{1})$, with $\alpha$ the
first extremum of $J_{n}$ (similar to a duct mode), and
$s_{n}'(r_{1})\equiv 1$. The wavenumber $k=1$ and values $n=2,8$ have
been used. In the first case, $n=2$, the noise reduction in the plane
$z=0$ is quite large, about 20\deci\bel, but there is a slight
increase around $z=1$. This is because, as seen in
Figure~\ref{fig:cfft:compare}, the coefficient $u_{2}$ is quite large
and is multiplied by a Bessel function of order~3, which is not of
high enough order for the exponential decay with $k$ which cuts off
the mode.

The cut-off behaviour is seen more clearly in the $n=8$ case, where
the reduction at $z=0$ is~40\deci\bel. At larger $z$, the reduction is
much smaller, but this is because, as found from asymptotic analysis
\citep{carley10b,carley10c}, the field in this region only contains
contributions from the remaining lower order modes, starting with
$q=2$.

\subsection{Degrees of freedom in jet noise fields}
\label{sec:jets}

The results of \S\,\ref{sec:radiated}, regarding the number of degrees
of freedom in the acoustic field, can help explain some features of
experiments on jet noise. Despite the lack of consensus on what is
meant by the `source' of jet noise
\citep[][]{jordan-gervais08,suzuki10}, some progress has been made by
assuming that the source of jet noise can be identified with some
combination of flow quantities. An open question, however, is which
part of the source term radiates, since it is clear that only a small
fraction of the flow generates the acoustic field. Two recent sets of
results, one experimental, the other numerical, illustrate the
issues. 

In one, \cite{freund01} has used direct numerical simulation to
compute the flow and noise of a Mach~0.9 jet, validating the noise
prediction against experiments and showing that a \cite{lighthill52}
source term accurately reproduces the acoustic field. Spatial
filtering of the source, using a wavenumber criterion to remove the
non-radiating part, left ``a set of modes capable of radiating to the
far field'', with the caveat that ``additional cancellation may occur
due to the radial structure of the source which is not accounted for
in this analysis''. Indeed, the radial structure of jet noise sources
has not received much attention until quite recently
\citep{michel09}. 

An experimental result of some interest is that of
\cite{jordan-schlegel-stalnov-noack-tinney07} who performed a modal
decomposition of a jet flow field and a proper orthogonal
decomposition optimized for the resolution of the far field
noise. They found that more than~350 modes were needed to capture half
of the flow energy while~24 modes sufficed for~90\% of the far-field
noise. As they note, passage to the far field acts as a filter passing
only a low-dimensional representation of the flow.

From these observations, it is plausible that the relatively low order
structure of jet noise can be explained by the results of this
paper. In the notation of this paper, $k=\upi \St M$, where $\St$ is
Strouhal number based on jet diameter and $M$ is jet Mach number. For
the range of Strouhal number important for jet noise $\St<2$
\citep{michalke-fuchs75}, $k<2\upi M$. For the $M=0.9$ jet studied by
\cite{jordan-schlegel-stalnov-noack-tinney07}, for example, this
yields $k\lesssim5.7$ and no more than about six line source modes
radiate from the axisymmetric source modes at the highest frequency of
interest. This estimate would need to be modified to take account of
axial interference as in the far field analysis of \cite{michel09} but
does offer the possibility of establishing some reasonable limits on
the detail to be expected from acoustic measurements on jets and the
requirements for low order models used in noise control.

\section{Conclusions}
\label{sec:conclusions}

% An analysis has been presented for the information content of a disc
% source with arbitrary radial source variation. This places a limit on
% the number of degrees of freedom in the acoustic field. On the basis
% of this general analysis, certain features of the acoustic field of
% rotors and jets have been explained. It has been demonstrated that the
% low order structure of the acoustic field of a turbulent jet may be a
% consequence of the radiation properties of circular sources and the
% possibility exists that the radiating part of a circular source may be
% explicitly identified. Further work will examine experimental and
% numerical data in the framework of the results of this paper.

The radiation properties of disc sources of arbitrary radial variation
have been analyzed to establish the part of the source which radiates
into the acoustic field, without recourse to a far field
approximation. Limits have been established on the number of degrees
of freedom of the part of the source which radiates and the
implications of these limits have been discussed for the problems of
rotor noise and studies of source mechanisms in jets. Future work will
consider the use of the findings of this paper to study the radiating
portion of full jet source distributions, including axial interference
effects.

%\bibliographystyle{jfm}

%\bibliography{short-abbrev,propnoise,misc,maths,identification,%
%  turbulence,jets}

\end{document}